\documentclass[aps,prc,showpacs,showkeys,tightenlines]{revtex4}

\usepackage{graphicx}

\newcommand{\be}{\begin{equation}}
\newcommand{\ee}{\end{equation}}
\newcommand{\bea}{\begin{eqnarray}}
\newcommand{\eea}{\end{eqnarray}}

\begin{document}

\title{Neutrinoless double $\beta$-decay and neutrino mass hierarchies}

\date{\today}

\author{S.M. Bilenky}
\email{bilenky@he.sissa.it}
\affiliation{Scuola Internazionale Superiore di Studi Avanzati, 
I-34014 Trieste, Italy}
\altaffiliation{On  leave of absence from 
the Joint Institute for Nuclear Research, 
141980 Dubna (Moscow Region), Russia} 
\author{Amand Faessler}
\email{amand.faessler@uni-tuebingen.de}
\affiliation{Institute f\"{u}r Theoretische Physik der Universit\"{a}t
T\"{u}bingen, D-72076 T\"{u}bingen, Germany}
\author{Thomas Gutsche}
\email{thomas.gutsche@uni-tuebingen.de}
\affiliation{Institute f\"{u}r Theoretische Physik der Universit\"{a}t
T\"{u}bingen, D-72076 T\"{u}bingen, Germany}
\author{Fedor \v Simkovic}
\email{fedor.simkovic@fmph.uniba.sk}
\altaffiliation{On  leave of absence from Department of Nuclear
Physics, Comenius University, Mlynsk\'a dolina F1, SK--842 15
Bratislava, Slovakia} 
\affiliation{Institute f\"{u}r Theoretische Physik der Universit\"{a}t
T\"{u}bingen, D-72076 T\"{u}bingen, Germany}

\begin{abstract}
In the framework of the see-saw mechanism the normal hierarchy 
is favorable for the neutrino mass spectrum. For this spectrum we present 
a detailed calculation of the  half-lives of neutrinoless double 
$\beta$-decay for several nuclei of experimental interest.
The half-lives are evaluated by considering the most comprehensive 
nuclear matrix elements, which were obtained within the renormalized QRPA by 
the Bratislava--Caltech--Tuebingen group. The dependence of the half-lives on 
$\sin^{2}\theta_{13}$ and the lightest neutrino mass is studied. We present 
also the results of the calculations of the  half-lives of neutrinoless double $\beta$-decay 
in  the case of the inverted hierarchy of neutrino masses. 
\end{abstract} 



\maketitle

\section{Introduction}

The discovery of neutrino oscillations in the atmospheric Super-Kamiokande experiment \cite{SK}, 
in the solar SNO experiment \cite{SNO}, 
in the reactor KamLAND experiment \cite{Kamland} in the accelerator K2K experiment 
\cite{K2K} and other neutrino experiments 
\cite{Cl,Gallex,Sage,SKsol} is one of the most compelling evidence in favor of new 
physics beyond the Standard Model. 
All existing neutrino oscillation data, with the exception of the data of 
the short baseline accelerator 
experiment LSND  \cite{LSND} (the LSND result will 
be checked by the running MiniBooNE experiment \cite{MiniB} soon)  
are described by the three-neutrino mixing scheme
\be
\nu_{l L}(x) = \sum_{i=1}^{3} U_{{l}i} ~ \nu_{iL}(x);~~l=e, \mu, \tau.
\label{eq:1}
\ee
Here, $\nu_{i}(x)$ is the field of the neutrino with mass 
$m_{i}$ ($i=1, 2, 3$) and $\nu_{l L}(x)$ is a flavor neutrino field which enters into the 
standard charged and neutral currents  
\be 
j^{CC}_{\alpha}(x) =2\, \sum_{l}\bar\nu_{l L}(x) \,\gamma_{\alpha}\,l_{L}(x), ~~~
j^{NC}_{\alpha}(x) = \sum_{l}\bar \nu_{l L}(x) \,\gamma_{\alpha}\,\nu_{l L}(x),
\label{eq:2}
\ee
$U$ is the unitary Pontecorvo-Maki-Nakagawa-Sakata (PMNS) \cite{BP,MNS} mixing matrix.
For massive Dirac neutrinos the PMNS matrix $U^D$ 
in the standard parameterization 
has the form
\begin{eqnarray}
U^D  =   \left (
  \begin{array}{ccc}
    c_{12} c_{13} & s_{12} c_{13} & s_{13} e^{-i \delta} \\
    -s_{12} c_{23} - c_{12} s_{23} s_{13} e^{i \delta} & c_{12} c_{23} - s_{12}
    s_{23} s_{13}e^{i \delta} & s_{23} c_{13} \\
    s_{12} s_{23} - c_{12} c_{23} s_{13}e^{i \delta} & -c_{12} s_{23} - s_{12}
    c_{23} s_{13}e^{i \delta} & c_{23} c_{13}
  \end{array}
  \right ). 
\label{eq:3}
\end{eqnarray}
Here $s_{ij} = \sin\theta_{ij}$, $c_{ij} = \cos\theta_{ij}$, 
$\theta_{ij}$ ($i<j$) is the neutrino mixing angle and $\delta$ is the CP violating phase.

From the analysis of the Super-Kamiokande atmospheric neutrino data for 
the neutrino mass squared difference $\Delta m^{2}_{\rm{23}}$ and the
parameter $\sin^2 2\theta_{\rm{23}}$ it was obtained  \cite{SK}: 
\begin{eqnarray}
{\rm{best~fit:}} &&\Delta m^{2}_{\rm{23}}= 2.1\,~10^{-3}\,~ \rm{eV}^{2}, ~~~
\sin^2 2\theta_{\rm{23}} = 1.00, \nonumber\\
{\rm{90\%~C.L.:}} &&  1.5\,~10^{-3}\,~\rm{eV}^{2}~\le  \Delta m^{2}_{\rm{23}}~\le 
3.4\,~10^{-3}\,~\rm{eV}^{2}, ~~\sin^2 2\theta_{\rm{23}} >  ~ 0.92. \
\label{eq:4}
\end{eqnarray}
The global analysis of the data of the solar neutrino experiments and  KamLAND experiment 
yields the following best fit values and 90\% C.L. ranges of the relevant neutrino oscillation parameters 
\cite{Kamland}: 
\begin{eqnarray}
{\rm{best~fit:}}~~~&&\Delta m^{2}_{{12}}
= 7.9\,~10^{-5}\,~ \rm{eV}^{2}, ~~~ tan^2 \theta_{{12}} = 0.40, \nonumber\\
{\rm{90\%~C.L.:}} &&  7.4\,~10^{-5}\,~\rm{eV}^{2}~\le \Delta m^{2}_{{12}}~\le 
8.5\,~10^{-5}\,~\rm{eV}^{2}, ~~ 0.33 ~\le   \tan^2 \theta_{{12}} \le~ 0.50. 
\label{eq:5}
\end{eqnarray}
Notice that neutrino mass-squared difference is determined as $\Delta m^{2}_{ik}=
 m^{2}_{k}- m^{2}_{i}$. For the angle $\theta_{13}$ only upper bound is known. From the
exclusion plot obtained from the data of the reactor experiment CHOOZ \cite{CHOOZ,fogli}
we have
\begin{equation}
\sin^{2} \theta_{13} \leq 5~10^{-2}~~~(90\%~C.L.).
\label{eq:6}
\end{equation}
The CP-violating phase $\delta$ remains undetermined. 
A recent global analysis of the oscillation data lead to the following bound:
$\sin^{2} \theta_{13} \leq 0.9^{+2.3}_{-0.9} 10^{-2}$ $(95\%~C.L.)$
\cite{global}.

At present the structure of the neutrino mass spectrum is not known as well. 
Two types of spectra are possible:
\begin{enumerate}
\item 
Normal spectrum:
\be
m_{1}< m_{2}<m_{3};~~\Delta m^{2}_{12}\ll\Delta m^{2}_{23}.
\label{eq:7}
\ee
\item
Inverted spectrum:
\be
m_{3}<m_{1} <m_{2};~~\Delta m^{2}_{12}\ll|\Delta m^{2}_{13}|
\label{eq:8}
\ee
\end{enumerate}
We note that it is common to label neutrino masses differently in the case of the 
normal and the inverted spectra. For both spectra we have $m_{2}>m_{1}$. But in the case 
of the normal spectrum $m_{3}$ is the mass of the heaviest neutrino and in the case 
of the inverted hierarchy $m_{3}$ is the mass of the lightest neutrino. This convention 
allows to keep the same notation of the mixing angles for both spectra.
Existing oscillation data are compatible both with normal and the inverted 
spectra.

The lightest neutrino mass $m_0=m_{1} (m_{3})$, which determines the absolute values of 
neutrino masses, is currently also unknown. From an analysis of the data of the 
Mainz \cite{mainz} and Troitsk \cite{troitsk} tritium  experiments it was found 
\be
m_{0}\leq 2.3 ~\rm{eV}.
\label{eq:9}
\ee

A more stringent  bound on the sum of neutrino masses can be found  from 
the measurement of the matter power spectrum $P(k)$.
Depending on the data which were taken into account, the cosmological upper bound
on the sum of neutrino masses was obtained as
(see \cite{cosmo1,Tegmark} and references therein)
\be
\sum_{i} m_{i} \leq (0.5-1.7)~\rm{eV}.
\label{9}
\ee

An important evidence that masses and mixing of neutrinos are of a nature 
beyond the Standard model (SM) would be 
that massive neutrinos are Majorana particles. If $\nu_{i}$ are Majorana particles
\begin{enumerate}
\item 
Neutrino fields $\nu_{i}(x)$ satisfy the Majorana conditions
\be
\nu^{c}_{i}(x)=\nu_{i}(x),
\label{eq:11}
\ee
where $\nu^{c}_{i}(x)=C\,\bar\nu^{T}_{i}(x)$ is the conjugated field 
($C$ is the charge conjugation matrix).
\item
The neutrino mixing matrix has the form \cite{BHP}
\be
U=U^{D}\,S(\alpha)
\label{eq:12}
\ee
where $S(\alpha)$ is a diagonal phase matrix. In the case of three
neutrino mixing the matrix   $S(\alpha)$ is characterized by two 
Majorana CP-violating phases. The matrix 
$S(\alpha)$ can be presented in the form 
\be
S_{ik}=e^{i\alpha_{i}}\,\delta_{ik};\,~ ~\alpha_{3}=0.
\label{eq:13}
\ee
The unitary matrix  $U^{D}$, which is characterized by the three mixing angles 
$\theta_{12}$, $\theta_{23}$, $\theta_{13}$ and one phase $\delta$, was 
already introduced in Eq. (\ref{eq:3}).
 \end{enumerate}
If in the lepton sector CP invariance holds, for the Majorana mixing matrix we have
\cite{BNP}
\be
U_{li}= U^{*}_{li}\,\eta_{i},
\label{eq:14}
\ee
where $\eta_{i}=\pm i$ is the CP parity of the Majorana neutrino $\nu_{i}$. The condition
(\ref{eq:14}) can be presented in the form
\be
U^{2}_{li}=|U_{li}|^{2}\,e^{i\,(\pi/2) \,\rho_{i}},
\label{eq:15}
\ee
where $\rho_{i}=\pm 1$.

Investigations of neutrino oscillations in vacuum and in matter do not allow to 
distinguish massive Dirac from massive Majorana neutrinos \cite{BHP,valle2,Langacker}. 
In order to reveal the Majorana nature of $\nu_{i}$ it 
is necessary to study processes in which the total lepton number is violated. 
Because the standard electroweak interaction conserves helicity the
probabilities of such processes  are proportional to the squares 
of the neutrino masses, and, consequently, they are strongly suppressed. 
The best sensitivity on small Majorana neutrino masses can be reached 
in the investigation of neutrinoless double $\beta$-decay 
($0\nu\beta\beta$) of some even-even nuclei.

\section{Neutrinoless double $\beta$-decay}

In the case of Majorana neutrino mixing and the standard electroweak CC interaction 
the $0\nu\beta\beta$-decay, 
\be
(A,Z) \to (A,Z+2) +e^{-}+ e^{-},
\label{eq:16}
\ee
is a second order process  in the Fermi constant $G_{F}$ with virtual neutrinos.
The half-life of the process is given by the following general expression 
\cite{dbdreviews}
\be 
\frac{1}{T^{0\nu}_{1/2}(A,Z)}=
|m_{\beta\beta}|^{2}\,|M^{0\nu}(A,Z)|^{2}\,G^{0\nu}(E_{0},Z).
\label{eq:17} 
\ee
Here
\be 
m_{\beta\beta}=
\sum_{i}U^{2}_{ei}\,m_{i}
\label{eq:18}
\ee 
is the 
effective Majorana mass, 
$M^{0\nu}(A,Z)$ is the nuclear matrix element (NME) and
$G^{0\nu}(E_{0},Z)$ is a known phase-space factor ($E_{0}$ is the energy release).
Let us stress that the NME is determined only by nuclear properties (its
dependence on the small neutrino masses can be safely neglected).

After the discovery of neutrino oscillations, the search for neutrinoless double beta decay 
($0\nu\beta\beta$-decay) became one of the most fundamental problems of neutrino physics. 
Observation of this process would be the proof that massive neutrinos $\nu_{i}$
are Majorana particles \cite{valle}. Furthermore the observation of 
$0\nu\beta\beta$-decay will allow to reveal the type of the neutrino mass spectrum,
to determine the mass of the lightest neutrino and, possibly, Majorana  CP phases.

The most stringent lower bounds on the half-life of $0\nu\beta\beta$-decay were obtained 
in the Heidelberg-Moscow $^{76} \rm{Ge}$  \cite{hm} and CUORICINO $^{130} \rm{Te}$ \cite{cino}
experiments:
\be 
T^{0\nu}_{1/2}(^{76} \rm{Ge})\geq 1.9 \cdot 10^{25}\, years,~~~~
T^{0\nu}_{1/2}(^{130} \rm{Te})\geq 1.8 \cdot 10^{24}\, years.
\label{eq:19}
\ee
Using recently calculated nuclear matrix elements with 
significantly reduced theoretical uncertainties \cite{rodin1,rodin2} from these data
the following  upper bounds  for the effective Majorana mass  can be inferred
\bea 
|m_{\beta\beta}| \leq
0.55 \,~\rm{eV}\,~~~(\rm{Heidelberg-Moscow})\nonumber\\
 |m_{\beta\beta}| \leq
1.1 \,~\rm{eV}\,~~~(\rm{CUORICINO}).
\label{eq:20} 
\eea
The Heidelberg group,which includes a few authors of the Heidelberg-Moscow  collaboration, 
recently claimed \cite{claim}  evidence for the $0\nu\beta\beta$-decay 
of $^{76} \rm{Ge}$ with 
$ T^{0\nu}_{1/2} = (0.69-4.18)\cdot 10^{25}$ years at the $4.2\sigma$ confidence 
level. Using the NME obtained in Ref. \cite{rodin1,rodin2}, 
from this data one finds for the effective Majorana mass  
the range $0.37~eV\le |m_{\beta\beta}| \le 0.91~eV$. 
The claim made in  \cite{claim} was  re-examined and critized by different authors \cite{crit} and 
in particular by  the Moscow participants of the Heidelberg-Moscow collaboration \cite{moscow}.
The GERDA I experiment  \cite{gerda}, now at preparation at Gran Sasso, will be able to check
relatively soon the claim made in  \cite{claim}.

The effective Majorana mass given by Eq.(\ref{eq:18}) is determined by the values 
of the neutrino masses $m_{i}$, which for the case of the normal  neutrino mass spectrum 
are given by 
\be
m_{2}=\sqrt{m^{2}_{1}+\Delta m^{2}_{12}},~~m_{3}=\sqrt{m^{2}_{1}+\Delta m^{2}_{12}+\Delta m^{2}_{23}},
\label{eq:21} 
\ee
 and by the matrix elements $U^{2}_{ei}$, which in the standard parameterization take the form
\be 
U^{2}_{e1}=\cos^{2}\theta_{13}\, \cos^{2}\theta_{12}\, e^{2i\alpha_{1}},~
U^{2}_{e2}=\cos^{2}\theta_{13}\, \sin^{2}\theta_{12}\, e^{2i\alpha_{2}},~
U^{2}_{e3}=\sin^{2}\theta_{13}\, e^{2i\alpha_{3}},
\label{eq:22} 
\ee
where $\alpha_{i}$ are Majorana phases.

The values of the neutrino masses depend on the lightest neutrino mass 
$m_0=m_{1}(m_{3})$, on
the neutrino mass spectrum and the neutrino mass squared differences $\Delta m^{2}_{12}$ 
and $\Delta m^{2}_{23}$ ( $|\Delta m^{2}_{13}|$),  which are known from neutrino oscillation data (see (\ref{eq:4}) and  (\ref{eq:5}) ).
The value of the parameter $\sin^{2}\theta_{12}$ is deduced from the analysis of the solar and KamLAND 
data (see (\ref{eq:5})). 
The lightest neutrino mass $m_0$ and the CP Majorana phases $\alpha_{i}$ 
are unknown and will be considered as free parameters. 

In the case of the normal mass hierarchy,
\be 
m_{1} \ll m_{2} \ll m_{3},
\label{eq:23}
\ee 
the lowest two neutrino masses and the effective Majorana mass $|m_{\beta\beta}|$ 
have the minimal values. 
If we neglect the contribution of  $m_{1}$ to (\ref{eq:18}),  for the upper bound of
$|m_{\beta\beta}|$ we get 
\be 
|m_{\beta\beta}| \leq (\sin^{2}\theta_{12}\,\sqrt{\Delta m^{2}_{12}}+ \sin^{2}\theta_{13}\,
\sqrt{\Delta m^{2}_{23}}).
\label{eq:24}
\ee 
The contribution of the first term to $|m_{\beta\beta}|$ is small because of 
the smallness of $\sqrt{\Delta m^{2}_{12}}$.
The contribution proportional to the ``large'' $\sqrt{\Delta m^{2}_{23}}$ 
is suppressed by the smallness of the parameter $\sin^{2}\theta_{13}$. 
With (\ref{eq:5}),  (\ref{eq:6}) and  (\ref{eq:24}) for the upper bound of 
the effective Majorana mass we find	
\be
|m_{\beta\beta}| \lesssim
6.4\cdot 10^{-3}.
\label{eq:25}
\ee 

In the case of the inverted hierarchy
\be
m_{3} \ll m_{1} <m_{2},
\label{eq:26}
\ee 
we can safely neglect the contribution of the lightest mass $m_{3}$ to the effective Majorana mass.
For $|m_{\beta\beta}| $ we have the following expression
\be 
|m_{\beta\beta}|\simeq \sqrt{|\Delta m^{2}_{13}|}\,~ (1-\sin^{2}
2\,\theta_{12}\,\sin^{2}\alpha_{21})^{1/2},
\label{eq:27}
\ee
where $\alpha_{21}=\alpha_{2}-\alpha_{1}$.
The only unknown parameter in (\ref{eq:27}) is $\sin^{2}\alpha_{12}$. From (\ref{eq:27}) we have
\be 
\sqrt{|\Delta m^{2}_{13}|} \,\cos  2\,\theta_{12}  \leq  |m_{\beta\beta}|\leq \sqrt{|\Delta m^{2}_{13}|}.
\label{eq:28}
\ee
The bounds in  (\ref{eq:28}) correspond to the case of the CP invariance in the lepton sector 
(the upper bound corresponds to 
$\rho_{1}=\rho_{2}$ and the lower bound corresponds to $\rho_{1}=-\rho_{2}$.) From 
(\ref{eq:4}),  (\ref{eq:5}) and (\ref{eq:28}) we find that in the case of the inverted hierarchy 
the value of $|m_{\beta\beta}|$ must lie  in the range
\be 
1.0\cdot 10^{-2}\leq |m_{\beta\beta}|\leq
5.5\cdot 10^{-2}~\rm{eV},
\label{eq:29}
\ee 
In the case of the quasi-degenerate spectrum of neutrino masses,
\be 
m_{1}<m_{2} <m_{3} 
;~~  m_{i} \simeq  m_{0}\gg \sqrt{ \Delta m^{2}_{23}},
\label{eq:30}
\ee
the effective Majorana mass is given by Eq. (\ref{eq:27}) in which the replacement 
$\sqrt{|\Delta m^{2}_{13}|} \to m_{0}$ must be performed. 

Thus, in the case of the quasi-degenerate spectrum the effective Majorana mass 
depends on two parameters: $ m_{0}$ and $\sin^{2}\alpha_{21}$. For the common neutrino mass 
we have 
\be 
 |m_{\beta\beta}|\leq m_{0} \leq \frac{|m_{\beta\beta}|}{\cos   2\,\theta_{12} }.
\label{eq:31}
\ee
The current bound on $ m_{0}$, obtained from the measurement of the high energy part of 
the $\beta$-spectrum of tritium, is given in (\ref{eq:9}). The future KATRIN 
tritium experiment \cite{Katrin} will be sensitive to a value of $m_{0}\simeq 2\cdot 10^{-1}$ eV.

There are many models of neutrino masses  (see e.g. Refs. \cite{Altarelli,Feruglio,Mohapatra}). 
The see-saw  mechanism of neutrino mass generation \cite{seesaw} is considered to be  the most plausible one.
This mechanism is based on the assumption that violation of the total lepton number is at a large scale 
and connects the smallness of the Majorana neutrino masses with heavy right-handed Majorana 
particles. The existence of such particles provide a
natural framework for the explanation of the baryon asymmetry of the 
Universe (see Ref. \cite{Buchmuller}).

In the framework of the see-saw mechanism the degenerate neutrino mass spectrum requires a fine-tuning 
which includes the Dirac mass matrix and the right-handed Majorana mass matrix.

The inverted hierarchy of neutrino masses requires a specific lepton symmetry like a 
global gauge symmetry which provide conservation of $L_{e}-L_{\mu}- L_{\tau}$. 
In order to explain existing data this symmetry must be broken. 
In such a framework it is difficult to reconcile the large $\theta_{23}$ with the not so 
maximal mixing angle  
$\theta_{12}$ and the small mixing angle $\theta_{13}$
\cite{Altarelli,Feruglio}. Neutrino mass hierarchy is a natural 
spectrum in the case of the see-saw mechanism. 
Such a spectrum is realized in the case of SO(10) \cite{moha10}
and another GUT models which connect quark and lepton sectors.
The problems of the CP phases and the renormalization group effects in the SO(10) GUT 
were discussed in \cite{fukuyama}.

In this paper we  calculate the half-lives of the $0\nu\beta\beta$-decay in the case of the normal and 
inverted hierarchies of neutrino masses. 
Neutrino masses for such spectra are very small.
For the case of the  normal hierarchy we have
\be
m_{2} \simeq 9\cdot 10^{-3}~\rm{eV};~~m_{3} \simeq 5\cdot 10^{-2}~\rm{eV};~~
m_{1}\ll m_{2}
\label{eq:32}
\ee
For the case of  the inverted hierarchy 
we obtain
\be
m_{1} \simeq m_{2} \simeq 5\cdot 10^{-2}~\rm{eV};~~
m_{3}\ll m_{1}.
\label{eq:33}
\ee
Effects of such small neutrino masses cannot be observed in tritium and other $\beta$-decay 
experiments in  a foreseeable future. Future data on the distribution of clusters of galaxies  and 
gravitational lensing data, however, will be sensitive to the following value of the sum of neutrino 
masses \cite{Wang,Tegmark}
\begin{equation}
\sum_{i}m_{i}~ \simeq ~3 \cdot 10^{-2}\rm{eV}.
\label{eq:34}
\end{equation} 
Thus, apparently, future cosmological measurements can probe neutrino mass hierarchies.

Several future experiments on the search for $0\nu\beta\beta$-decay
will be sensitive to values of the effective Majorana mass 
in the range (\ref{eq:29}),which corresponds to  
the inverted mass hierarchy (see Table \ref{tab.1}). 

\begin{table}[t]
  \begin{center}
    \caption{Sensitivities of future $0\nu\beta\beta$-decay experiments to the 
effective Majorana neutrino mass $|m_{\beta\beta}|$ 
calculated with 
the RQRPA nuclear matrix elements $M^{0\nu}(A,Z)$ of Ref. \protect\cite{rodin2}. 
For the axial coupling constant $g_A$ the value $g_A=1.25$ was assumed. 
$T^{0\nu-exp}_{1/2}$ is the
maximal half-life, which can be reached in the experiment and
$|m_{\beta\beta}|$ is the corresponding upper limit of the effective Majorana neutrino mass.
}
    \label{tab.1}
\renewcommand{\arraystretch}{1.2}
\begin{tabular}{lccccccc}
\hline\hline
Nucleus & Experiment & Source  & $T^{0\nu-exp}_{1/2}$ [yr] & Ref. & & $M^{0\nu}(A,Z)$ &
$|m_{\beta\beta}|$ [eV] \\ \hline
$^{76}Ge$ & GERDA(I)  & 15 kg of $^{enr}Ge$   & $3~ 10^{25}$ & \cite{gerda} &  & 2.40  & $0.44$ \\ 
          & GERDA(II)  & 100 kg of $^{enr}Ge$ & $2~ 10^{26}$ & \cite{gerda} &  & 2.40  & $0.17$ \\ 
          & Majorana & 0.5 t of $^{enr}Ge$ & $4~ 10^{27}$ & \cite{majorana} &  & 2.40  & $0.038$ \\
$^{82}Se$ & SuperNEMO & 100 kg of $^{enr}Se$ & $2~ 10^{26}$ & \cite{superN} &  & 2.12 & $0.091$ \\ 
$^{100}Mo$ & MOON  & 3.4 t of $^{nat}Mo$ & $1~ 10^{27}$ & \cite{EV2002} &  & 1.16 & $0.058$ \\ 
$^{116}Cd$ & CAMEO & 1 t of $CdWO_4$ crystals & $\approx 10^{26}$ & \cite{EV2002} &  & 1.43 & $0.14$ \\ 
$^{130}Te$ & CUORE & 750 kg of $TeO_2$  & $\approx 10^{27}$ & \cite{giuli} &  & 1.47  & $0.047$ \\ 
$^{136}Xe$ & XMASS & 10 t of liq. Xe & $3~ 10^{26}$ & \cite{EV2002} &  & 0.98 & $0.12$ \\ 
           &       &                 &                   &          &  & 0.73 & $0.17$ \\
           & EXO   & 1 t $^{enr}Xe$  & $2~ 10^{27}$ & \cite{gratta} &  & 0.98 & $0.048$ \\ 
           &       &                 &                   &          &  & 0.73 & $0.064$ \\
\hline\hline
\end{tabular}
  \end{center}
\end{table}

Taking into account the theoretical plausibility of the normal hierarchy of neutrino masses we believe 
that it is worthwhile to consider the $0\nu\beta\beta$-decay in detail for this type of
neutrino mass spectrum. Recently, important progress 
in the evaluation of the $0\nu\beta\beta$-decay NME's for the ground state transitions 
of nuclei of experimental interest was achieved \cite{rodin1,rodin2}. We shall use these new  results
to calculate expected half-lives of $0\nu\beta\beta$-decay 
for both neutrino mass hierarchies.

\section{Nuclear matrix elements}

From the measurement of the half-life of the $0\nu\beta\beta$-decay 
only the product $|m_{\beta\beta}|~|M^{0\nu}(A,Z)|$ can be determined. 
Thus, without accurate calculation of nuclear matrix elements, it
is not possible  to reach qualitative conclusions about neutrino masses
and the type of neutrino mass spectrum\cite{hier2,hier1,hier3,zralek,japan,bfs}.

The calculation of the $0\nu\beta\beta$-decay matrix elements is a difficult 
problem because  ground and many excited states of open-shell nuclei
with complicated nuclear structure have to be considered.
In the calculation  of the $0\nu\beta\beta$-decay  NME's the nuclear shell model 
(NSM) and the proton-neutron quasiparticle random phase approximation (pn-QRPA) or 
extensions to it \cite{engel} are used.

These two approaches are significantly different. The NSM is limited to a set of
single-particle states in the vicinity of the Fermi level. Thus, configurations
with only small excitations are considered.  These excitations  are correlated in
all possible ways. The open problem is the  effect of single-particle
states further away from the Fermi level, which is neglected.
The large NSM spaces  face the problem of diagonalization of large matrices 
and the construction of good effective interaction.

From the available  shell-model calculations of the NME of the $0\nu\beta\beta$-decay 
the most advanced are  those of the Strasbourg group \cite{caurier}, which
appeared about ten years ago. From that time, in spite of significant progress 
in computer speed and memory no large-scale NSM
calculations of the $0\nu\beta\beta$-decay NME have been published.

The pn-QRPA method takes into account many single-particle states
including states relatively far from the Fermi surface but with 
correlations which are of the
specific simple type. Many extensions of the pn-QRPA have been proposed
with the aim to improve the many-body approximation  scheme. 
The so-called renormalized pn-QRPA (pn-RQRPA) scheme takes into account the Pauli 
exclusion principle by improving the quasi-boson approximation \cite{rqrpa}. 
In this method for the evaluation of  the commutators of bifermion operators 
the QRPA ground state is used (instead of the uncorrelated 
BCS ground state). This refinement of QRPA 
approach allows to reduce  significantly the sensitivity  of the results 
to the renormalization of the particle-particle interaction of the nuclear Hamiltonian. 
The fact that pn-RQRPA is an improvement of the pn-QRPA approach
has been confirmed by the studies performed within 
a schematic model \cite{schem}.

The pn-QRPA and its extensions remain a popular technique of the calculation of
$0\nu\beta\beta$-decay NME's. However, various implementations of
the QRPA introduced by different authors have produced a spread of results with
a factor of 2-3 and even more difference in the  NME \cite{civi}.
Some authors simplified this problem by assuming that the published range 
of calculated NME's defines a plausible approximation to the uncertainty 
in our knowledge of the matrix elements \cite{civi,bahc}. We do not share 
this position. We believe that it is not appropriate to consider all calculated 
$0\nu\beta\beta$-decay NME's at the same level. 
The correct procedure is to understand the difference among various QRPA-like 
calculations and the origin of contradictions between different results.
In Ref. \cite{rodin2} a list of main 
reasons leading to a spread of the pn-QRPA  and the pn-RQRPA nuclear matrix elements 
was presented. 

One of the most important factors of the QRPA calculation of the
NME's  is the way how the particle-particle strength of the nuclear 
Hamiltonian $g_{pp}$  is fixed. When the early QRPA
calculations were performed only a limited information on half-lives 
of the $2\nu\beta\beta$-decay was available. At that time
$g_{pp}$ was fitted to existing data on single $\beta^-$-transitions
or alternatively $g_{pp}$ was chosen to be equal to unity. Nowadays,  
the half-lives of $2\nu\beta\beta$-decay of ten nuclei has been measured.
Recently,
it has been shown that by adjusting $g_{pp}$ to the 
$2\nu\beta\beta$-decay rates we can significantly eliminate uncertainties
associated with variations in QRPA calculations of decay rates \cite{rodin1,rodin2}. 
In particular, the results obtained in this way are essentially independent of the size
of the basis, the form of different realistic nucleon-nucleon
potentials, or on whether QRPA or RQRPA is used. Furthermore,
the matrix elements are shown to be also rather stable with respect
to the possible quenching of the axial vector constant $g_A$.

The procedure proposed in \cite{rodin1,rodin2} was  critically analyzed
in \cite{suhocr}.
 The author's 
conclusion was that fitting of $g_{pp}$ to $\beta^+$ (or electron capture (EC)) and 
single $\beta^-$-decay of the ground state of the intermediate
nucleus  is a more meaningful procedure.  This criticism has been refuted in Ref. \cite{rodin2}.   
In particular, it was shown that there is no reason to give 
preference to the lowest state of the intermediate nucleus. 
The $\beta^-$ and $\beta^+/EC$ matrix elements move with
$g_{pp}$ in opposite directions, what makes it difficult to
adjust $g_{pp}$ by choosing one of them. It is preferable
to use sum of the products of amplitudes, i.e., the 
$2\nu\beta\beta$-decay half-life. It was also noticed that practically for
all multipolarities significant amount of strength is concentrated
up to 10-15 MeV and that the contributions of the $1^+$
multipole to the $2\nu\beta\beta$- and $0\nu\beta\beta$-decay
matrix elements are correlated. Thus, there is no reason to choose 
any one particular state or transition for adjustment. 
It is worth also mentioning that only for three  double $\beta$-decay 
nuclear systems of interest  (A = 100, 116, 128) the 
ground state of the intermediate nucleus is just the $1^+$ state. Thus, 
the procedure of how to fix $g_{pp}$ proposed in  Ref. \cite{suhocr} is 
not only disfavored but also strongly limited. These and other arguments
of Ref. \cite{rodin2} clearly favor the procedure of the $2\nu\beta\beta$-decay 
fitting rather than the procedure of the beta-decay fitting. 
 
A discussion concerning the previous QRPA and RQRPA calculations of the
$0\nu\beta\beta$-decay NME's, in which 
different assumptions and approximations were made,
can be found in \cite{rodin2}. It was shown that in most, albeit not all,
cases the differences among them can be understood. Attention was also paid
to the fact that the results of \cite{stoica} differ significantly 
from those  of \cite{rodin1,rodin2} in spite of the fact that the same procedure of adjustment
of $g_{pp}$ was used (for the most important $1^+$ channel). Contrary to the NME
of \cite{rodin1,rodin2} the  NME calculated in  \cite{stoica} strongly depend  
on the size of the model space. In particular the levels lying far from the Fermi 
surface severely influence the  decay rate. Actually, there is no explanation of this fact 
and nobody else reported such a strong effect. 
 
The calculations performed in 
Ref. \cite{rodin2} can be considered as the most reliable 
QRPA and RQRPA calculations of the NME of the $0\nu\beta\beta$-decay.
We stress that the NME calculations in which the dependence of the results
 on the particular choice made is not discussed 
could  not be considered on the same footing as those where these points are 
carefully explained (see e.g. Ref. \cite{rodin2}).   

There are no doubts that further progress in the calculation of the 
$0\nu\beta\beta$-decay NME's is needed. The nuclei 
of experimental interest for the investigation of 
$0\nu\beta\beta$-decay
have been extensively studied by assuming spherical 
symmetry. However, many of the nuclei undergoing double beta decay are deformed 
\cite{pedro} and it is important  to study the effect of deformation on 
the $0\nu\beta\beta$-decay. Recently  a new suppression mechanism of the  
$2\nu\beta\beta$-decay matrix elements based on the relative deformation of the 
initial and final nuclei has been found \cite{deform}. 
The effect of deformation is large, if there is a significant difference 
in the deformations of the  parent and daughter nuclei. The effect of deformations
could also be large  in the case of the $0\nu\beta\beta$-decay matrix elements.

For further progress in the field it is also important to apply the methods of the 
calculation of the $0\nu\beta\beta$-decay matrix elements 
to the calculation of the matrix elements of related 
processes like charge-exchange reactions \cite{frekers}, muon capture \cite{muon}
and charged current (anti)neutrino-nucleus reactions \cite{crist}.
The observation of these processes might probe forbidden transitions which contribute
to the  $0\nu\beta\beta$-decay half-lives. Thus, it is essential to understand
how the  methods applied for the calculation of the NME of the
$0\nu\beta\beta$-decay can reproduce the observables related to
these processes. For the improvement of many-body approaches it is also
important to test them in the cases of exact solutions of solvable models
which are as  realistic as possible.  

The improvement of the calculation of the nuclear matrix elements is a very 
important and challenging problem. The nuclear matrix elements of the 
$0\nu\beta\beta$-decay cannot be related exactly to other observables. 
A complementary experimental information from related 
processes is highly required, but it cannot fully solve the problem of the
uncertainties in the $0\nu\beta\beta$-decay matrix elements. This problem might be
solved only by observation of the $0\nu\beta\beta$-decay of at least three
different nuclei \cite{bfs,test}. This would be a model independent test 
of the theoretically calculated NME's.  

\begin{figure}[t]
\includegraphics[width=12cm]{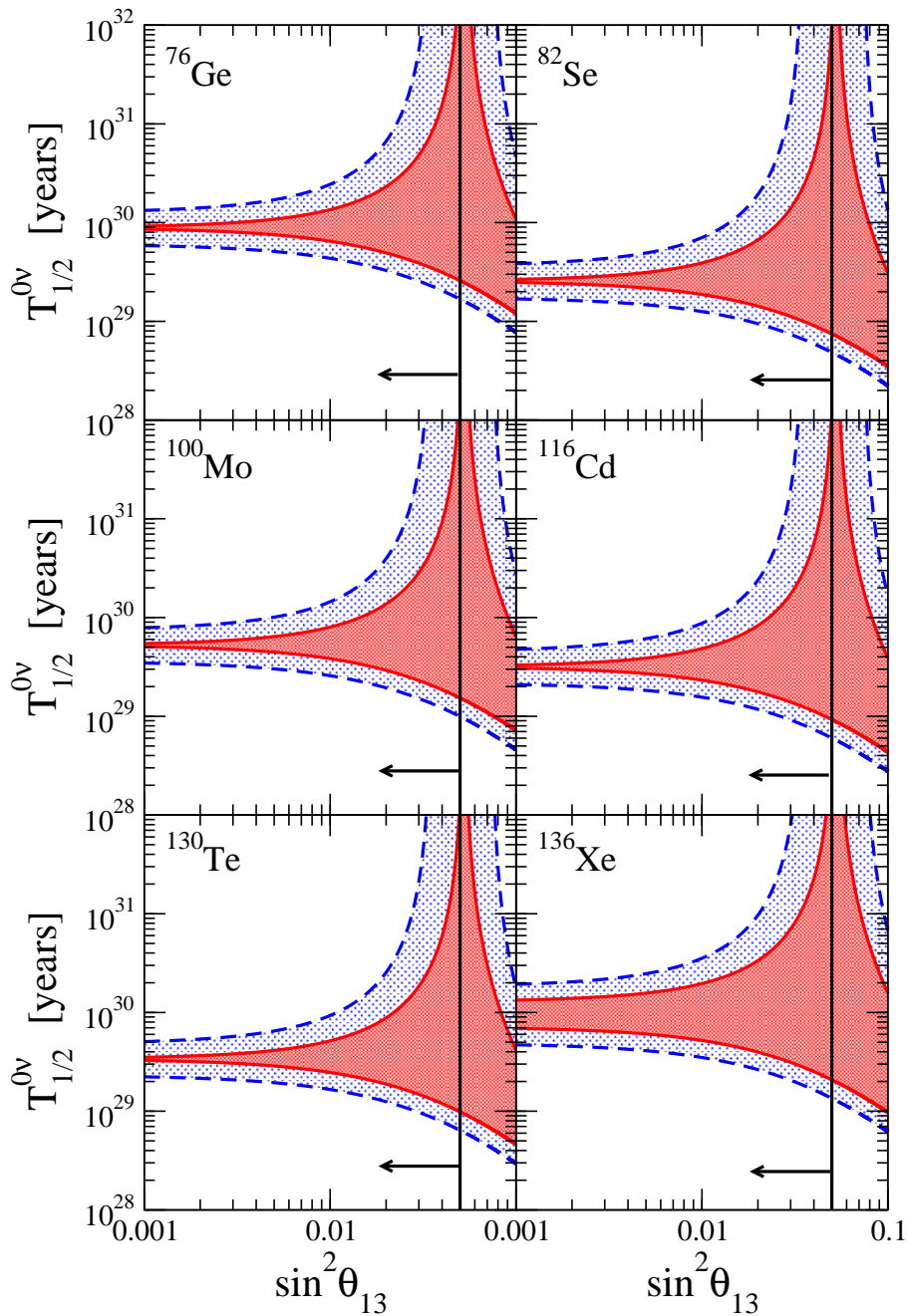}
    \caption{
The neutrinoless double beta decay half-life $T^{0\nu}_{1/2}$
for nuclei of experimental interest 
as function of the parameter $\sin^{2}\theta_{13}$.
The case of the normal hierarchy of neutrino masses is
considered and the lightest neutrino mass is assumed to be negligibly small.
The region with solid line (dashed line) boundaries 
corresponds to the best fit (90\% C.L.)of the  neutrino oscillation 
parameters  \protect\cite{SK,Kamland}. The calculations
were performed by using recently evaluated nuclear matrix elements with 
significantly reduced theoretical uncertainty \protect\cite{rodin2}. 
The vertical line indicates the current upper limit on $\sin^2\theta_{13}$
set by the CHOOZ experiment.
\label{fig.1}
}
\end{figure}

\section{The expected $0\nu\beta\beta$-decay half-lives}

\begin{table}[t]
  \begin{center}
    \caption{Normal hierarchy of neutrino masses: The neutrinoless double beta decay half-life 
$T^{0\nu}_{1/2}(A,Z)$ of 
$^{76} \rm{Ge}$,  
$^{82} \rm{Se}$, 
$^{96} \rm{Zr}$, 
$^{100} \rm{Mo}$, $^{116} \rm{Cd}$, $^{128} \rm{Te}$, $^{130} \rm{Te}$
 $^{136} \rm{Xe}$ and  $^{150} \rm{Nd}$
The results are presented for three values of 
$\theta_{13}$ from the allowed range $sin^2 \theta_{13} \le 0.05$ of CHOOZ
\protect\cite{fogli}. The best fit and 90\% C.L. values 
of $T^{0\nu}_{1/2}$  were calculated by assuming equal values of Majorana 
CP-violating phases, i.e., the lowest allowed values for $T^{0\nu}_{1/2}(A,Z)$ are
given. The $0\nu\beta\beta$-decay nuclear matrix elements of 
Ref. \protect\cite{rodin2} are used.  }
\label{tab.2}
\begin{tabular}{lcccccc}
\hline\hline
       &  &  \multicolumn{3}{c}{Normal hierarchy: $T^{0\nu}_{1/2}(A,Z)$ [years]} \\
 Nuclear & parameter  &  \multicolumn{3}{c}{$\sin^2 \theta_{13}$} \\ \cline{3-5}
transition  &  set & $0.05$ & $0.01$ & $0.001$ \\ \hline 
$^{76}Ge\rightarrow {^{76}Se}$ & best.fit  &
$2.6~10^{29}$ & $6.5~10^{29}$ & $8.6~10^{29}$ \\
                               & 90\% C.L. &
$1.7~10^{29}$ & $4.4~10^{29}$ & $5.8~10^{29}$ \\
$^{82}Se\rightarrow {^{82}Kr}$ & best fit &
$7.5~10^{28}$ & $1.9~10^{29}$ & $2.5~10^{29}$ \\
                               & 90\% C.L. &
$4.8~10^{28}$ & $1.3~10^{29}$ & $1.7~10^{29}$ \\
$^{96}Zr\rightarrow {^{96}Mo}$ & best.fit  &  
$1.7~10^{30}$ & $4.2~10^{30}$ & $5.5~10^{30}$ \\
                               & 90\% C.L. &
$1.1~10^{30}$ & $2.8~10^{30}$ & $3.8~10^{30}$ & \\ 
$^{100}Mo\rightarrow {^{100}Ru}$ & best.fit  &  
$1.5~10^{29}$ & $3.8~10^{29}$ & $5.1~10^{29}$ \\
                               & 90\% C.L. &
$9.9~10^{28}$ & $2.6~10^{29}$ & $3.5~10^{29}$ \\
$^{116}Cd\rightarrow {^{116}Sn}$  & best.fit  & 
$9.3~10^{28}$ & $2.3~10^{29}$ & $3.1~10^{29}$ \\
                               & 90\% C.L. &
$6.0~10^{28}$ & $1.6~10^{29}$ & $2.1~10^{29}$ \\
$^{128}Te\rightarrow {^{128}Xe}$  & best.fit  & 
$2.1~10^{30}$ & $5.2~10^{30}$ & $6.9~10^{30}$ \\
                               & 90\% C.L. &
$1.4~10^{30}$ & $3.5~10^{30}$ & $4.7~10^{30}$ \\
$^{130}Te\rightarrow {^{130}Xe}$  & best.fit  &
$9.9~10^{28}$ & $2.5~10^{29}$ & $3.3~10^{29}$ \\
                               & 90\% C.L. &
$6.4~10^{28}$ & $1.7~10^{29}$ & $2.2~10^{29}$ \\
$^{136}Xe\rightarrow {^{136}Ba}$ & best.fit  &  
$2.1~10^{29}$ & $5.2~10^{29}$ & $6.9~10^{29}$ \\
                               & 90\% C.L. &
$1.4~10^{29}$ & $3.5~10^{29}$ & $4.7~10^{29}$ \\
$^{150}Nd\rightarrow {^{150}Sm}$ & best.fit  &  
$1.0~10^{28}$ & $2.6~10^{28}$ & $3.5~10^{28}$ \\
                               & 90\% C.L. &
$6.8~10^{27}$ & $1.8~10^{28}$ & $2.4~10^{28}$ \\
\hline\hline
\end{tabular}
  \end{center}
\end{table}

Here, we present the results of the calculation of the  half-lives 
of the $0\nu\beta\beta$-decay of following nuclei 
 $^{76} \rm{Ge}$,  
$^{82} \rm{Se}$, $^{100} \rm{Mo}$, $^{116} \rm{Cd}$, $^{130} \rm{Te}$
and 
$^{136} \rm{Xe}$
many of which presumably 
will be investigated in future experiments.
We will use nuclear matrix elements $M^{0\nu}(A,Z)$ obtained in
Ref. \cite{rodin2}. 
%

First discuss the case of the normal hierarchy of neutrino masses. 
Taking into account all terms for $|m_{\beta\beta}|$ we have
the following expression
\be 
|m_{\beta\beta}| = |\cos^{2}\theta_{12}\,m_{1}+ \sin^{2}\theta_{12}\,\sqrt{\Delta m^{2}_{12}}\,e^{2i\alpha_{12}}
+ \sin^{2}\theta_{13}\,
\sqrt{\Delta m^{2}_{23}}\,e^{2i\alpha_{13}}|.
\label{eq:37}
\ee 
There are four unknown parameters in (\ref{eq:37}), namely $m_{1}$, $\sin^{2}\theta_{13}$ 
and two CP phase differences $\alpha_{12}$ and $\alpha_{13}$. First
we shall assume that the lightest mass  $m_{1}$ is very small and 
we can neglect its contribution to $|m_{\beta\beta}|$.
 In this case the effective 
Majorana mass $m_{\beta\beta}$ depends on two free parameters: $\sin^{2}\theta_{13}$ and $\alpha_{23}$.

\begin{figure}[t]
\includegraphics[width=13cm]{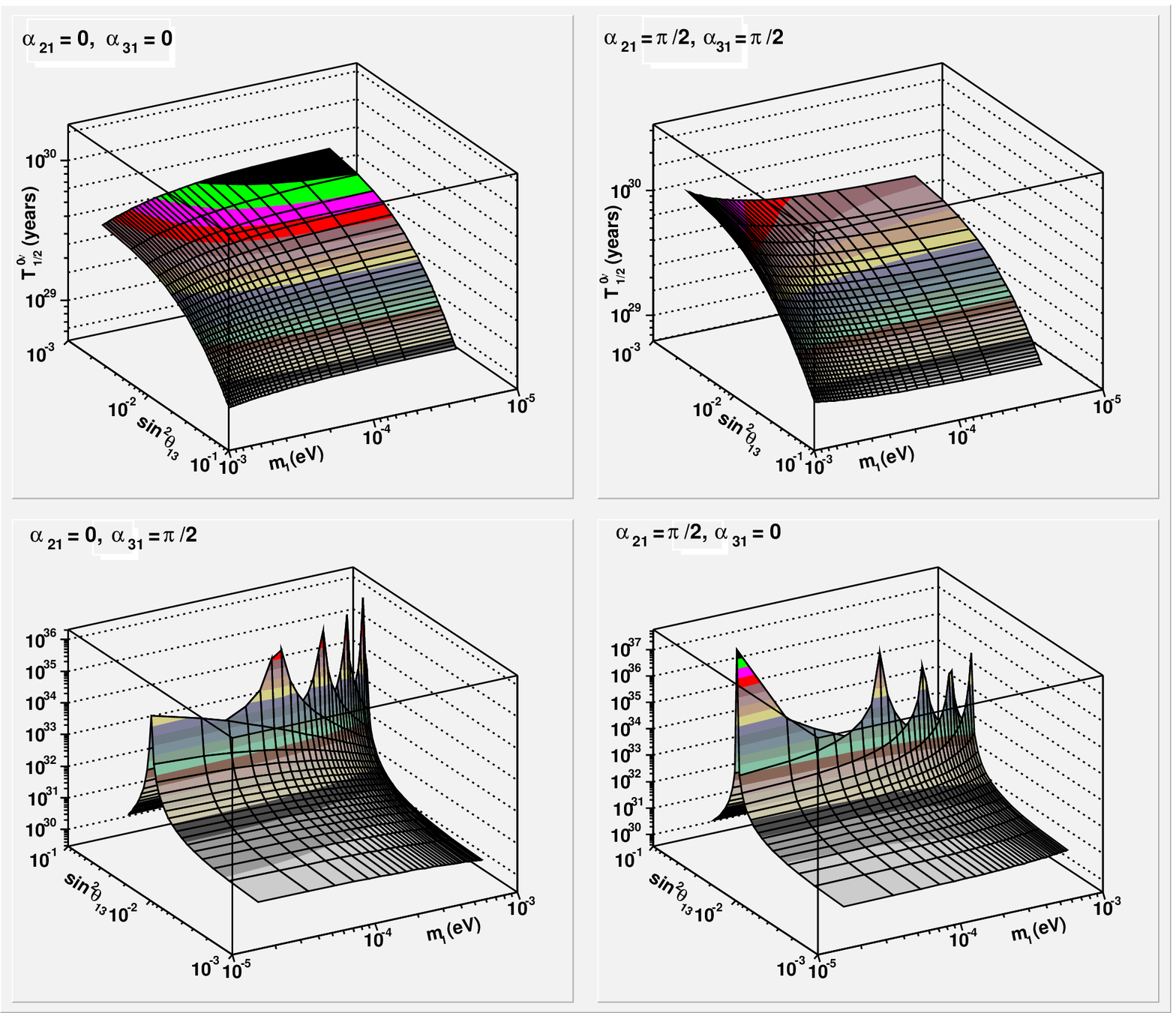}
    \caption{
The neutrinoless double beta half-life $T^{0\nu}_{1/2}$ of $^{76}Ge$
as a function of the parameter  $\sin^{2}\theta_{13}$
and the lightest neutrino mass $m_1$. The normal hierarchy of
neutrino masses is assumed and the best fit values of neutrino
oscillation parameters are taken into account. 
The relative CP Majorana phases of neutrinos are denoted by
$\alpha_{21}$ and $\alpha_{31}$.
\label{fig.2}
 }
\end{figure}

The current 90 \% CL CHOOZ bound on the parameter  
$\sin^{2}\theta_{13}$
is given in (\ref{eq:6}).
A significant improvement  on the value of this parameter
is planned to be achieved in future experiments.
The expected sensitivity of the Double-CHOOZ reactor experiment  \cite{Chooz2} is 
$\sin^{2}\theta_{13}\simeq 7.5\cdot 10^{-3}$. In the accelerator 
T2K $\nu_{\mu}\to\nu_{e} $ experiment \cite{T2K}
a sensitivity of $\sin^{2}\theta_{13}\simeq 1,5\cdot 10^{-3}$ 
is planned to be achieved.
Thus,
we will consider for the parameter $\sin^{2}\theta_{13}$ the range ($10^{-3}- 5\cdot 10^{-2}$). 

\begin{table}[t]
  \begin{center}
    \caption{
Inverted hierarchy of neutrino masses: The neutrinoless double beta decay half-life 
$T^{0\nu}_{1/2}(A,Z)$ calculated in RQRPA for 
$^{76} \rm{Ge}$,  
$^{82} \rm{Se}$, 
$^{96} \rm{Zr}$, 
$^{100} \rm{Mo}$, $^{116} \rm{Cd}$, $^{128} \rm{Te}$, $^{130} \rm{Te}$
 $^{136} \rm{Xe}$ and  $^{150} \rm{Nd}$.
$T^{0\nu}_{1/2}$ was calculated for best fit values and 90\% C.L. ranges of 
the corresponding parameters.
The $0\nu\beta\beta$-decay nuclear matrix elements of Ref. \protect\cite{rodin2} are used.  
The current experimental lower limits on the $0\nu\beta\beta$-decay half-lives 
($T^{0\nu-exp}_{1/2}$) are from experiments with saturated sensitivity except
those denoted by the symbol $^*$, which  indicate experiments still taking data.
H-M means Heidelberg-Moscow.}
\label{tab.3}
\renewcommand{\arraystretch}{1.2}
\begin{tabular}{lccccc}
\hline\hline
       &  \multicolumn{2}{c}{Inverted hierarchy: $T^{0\nu}_{1/2}(A,Z)$ [years]} & &
  \multicolumn{2}{c}{$T^{0\nu-exp}_{1/2}(A,Z)$ [years]} \\ \cline{2-3}  \cline{5-6}
 Nucleus &  \multicolumn{2}{c}{parameter set} & & current &  planed sensitivity \\ 
     & 90\% C.L.& best fit & & limit & of future exper. \\ \hline 
$^{76}Ge$ & 
 $(1.7~10^{27}, 4.1~10^{28})$ & $(2.8~10^{27}, 1.8~10^{28})$  & &
$1.9~10^{25}$ H-M \cite{hm} & $3~ 10^{25}$ GERDA I \cite{gerda}  \\ 
 & & & &  $1.6~10^{25}$ IGEX \cite{igex} & $2~10^{26}$  GERDA II \cite{gerda}  \\ 
 & & & & & $4~10^{27}$ Majorana \cite{majorana} \\
$^{82}Se$ &
$(4.9~10^{26}, 1.2~10^{28})$ &  $(8.1~10^{26}, 5.3~10^{27})$ & &
 $1.0~10^{23}$ NEMO $3^*$ \cite{nemo3} & $2~10^{26}$ SuperNEMO \cite{superN} \\
$^{96}Zr$ &
$(1.1~10^{28}, 2.6~10^{29})$ &  $(1.8~10^{28}, 1.2~10^{29})$  & &
  $1~10^{21}$ NEMO 2 \cite{zr96} & \\
$^{100}Mo$ &
$(1.0~10^{27}, 2.4~10^{28})$ & $(1.7~10^{27}, 1.1~10^{28})$  & &
 $4.6~10^{23}$ NEMO $3^*$ \cite{nemo3} & $1~10^{27}$ MOON \cite{EV2002} \\
$^{116}Cd$ &
$(6.1~10^{26}, 1.5~10^{28})$ &  $(1.0~10^{27}, 6.5~10^{27})$  & &
 $1.7~10^{23}$ \cite{cd116} & $\approx 10^{26}$ CAMEO \cite{superN}\\
$^{128}Te$ & 
$(1.4~10^{28}, 3.3~10^{29})$ & $(2.3~10^{28}, 1.5~10^{29})$  & &
$2~10^{24}$ \cite{te128} & \\
$^{130}Te$ & 
$(6.5~10^{26}, 1.6~10^{28})$ & $(1.1~10^{27}, 7.0~10^{27})$  & &
$1.8~10^{24}$ Cuoricino$^*$ \cite{cino} & $\approx 10^{27}$ CUORE \cite{giuli}\\
$^{136}Xe$ &
$(1.4~10^{27}, 5.9~10^{28})$ & $(2.3~10^{27}, 2.7~10^{28})$  & &
$1.2~10^{24}$ DAMA \cite{dama} & $2~10^{27}$ EXO \cite{gratta} \\
 & & & & & $3\times 10^{26}$ \cite{EV2002} \\
$^{150}Nd$ & 
$(6.9~10^{25}, 1.7~10^{27})$ & $(1.1~10^{26}, 7.4~10^{26})$  & &
$1.2~10^{21}$ \cite{nd150}
 & \\
\hline\hline
\end{tabular}
  \end{center}
\end{table}

In Fig. \ref{fig.1} we present the half-life $T^{0\nu}_{1/2}(A,Z)$ as a function of  
$\sin^{2}\theta_{13}$ for the nuclei $^{76} \rm{Ge}$,  
$^{82} \rm{Se}$, $^{100} \rm{Mo}$, $^{116} \rm{Cd}$, $^{130} \rm{Te}$
and 
$^{136} \rm{Xe}$.
The regions  with solid line
boundaries (dashed line boundaries) were calculated with the best fit values ( $90\%~C.L.$ values)
of the neutrino oscillation parameters $\Delta m^{2}_{23}$, $\Delta m^{2}_{12}$ 
and $\sin^{2}\theta_{12}$ (see Eqs. (\ref{eq:4}) and (\ref{eq:5})).
The boundaries of these regions correspond to the case of CP-conservation.
The lower (upper) boundaries correspond to $\cos 2\alpha_{23} = 1$ 
($\cos 2\alpha_{23} = - 1$), i.e., to the case of same 
(opposite) CP parities of the second and third neutrino.
For the case of the $0\nu\beta\beta$-decay of $^{136}Xe$
the allowed regions in Fig.\ref{fig.1} are larger 
than for other nuclei.This is connected to the fact 
that the corresponding NME is constrained
to a range of values since the $2\nu\beta\beta$-decay of this
isotope has not been measured yet \cite{rodin2}.

From Fig. \ref{fig.1} it follows that for $\sin^{2}\theta_{13}\leq 10^{-2}$ 
the calculated half-lives practically do not depend on $\sin^{2}\theta_{13}$. 
For such small values of $\sin^{2}\theta_{13}$ 
the dominant contribution to the effective Majorana mass $m_{\beta\beta}$ 
comes from the solar term (the second term on the right hand side of Eq.(\ref{eq:37})).   
From Fig. \ref{fig.1} we also deduce that if the limit on $\sin^{2}\theta_{13}$ 
will be pushed further down by the Double-CHOOZ and T2K experiments the
expected half-lives of the $0\nu\beta\beta$-decay of $^{82} \rm{Se}$, $^{100} \rm{Mo}$, 
$^{116} \rm{Cd}$, $^{130} \rm{Te}$ and  $^{136} \rm{Xe}$ will be about
$10^{29}-10^{30}$ years.

In the Table \ref{tab.2} we present the minimal values of $T^{0\nu}_{1/2}(A,Z)$ 
in the case of the neutrino mass hierarchy with a negligibly small lightest mass.
They are given for three possible values of the parameter $\sin^{2}\theta_{13}$:
$0.05,~0.01,~0.001$. These values are expected to be slightly
increased  if the limit on the parameter $\sin^{2}\theta_{13}$ 
will further decrease (see Fig. \ref{fig.1}). These values can be confronted with the 
sensitivities on half-lives of future $0\nu\beta\beta$-decay experiments
given in Table \ref{tab.1}. From the comparison of the calculated half-lives with
the experimental sensitivities  we conclude that further experimental
efforts will be required to 
reach lower limits comparable to the predicted values in the case  of the
normal hierarchy of neutrino masses.

\begin{figure}[t]
\includegraphics[width=12cm]{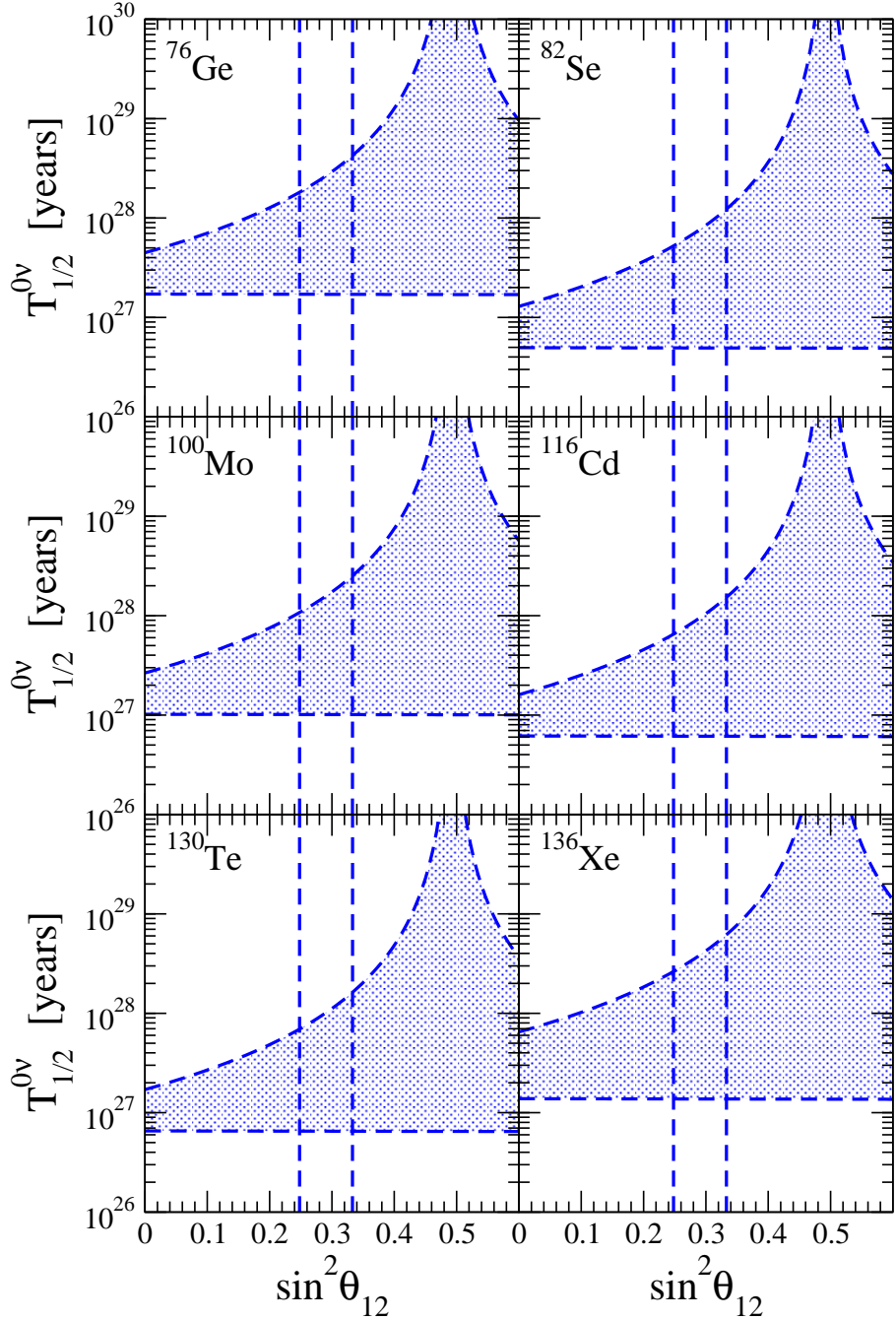}
    \caption{
The neutrinoless double beta decay half-life $T^{0\nu}_{1/2}$
for nuclei of experimental interest 
as function of the parameter $\sin^{2}\theta_{12}$.
The case of the inverted hierarchy of neutrino masses is
assumed. Conventions are the same as in 
Fig. \protect\ref{fig.1}.The vertical dashed lines 
correspond to the boundaries set by the 90\% C.L. values
of $\theta_{12}$ \protect\cite{Kamland}.
These results are not sensitive to $m_3$ for values below $10^{-2}~eV$
\protect\cite{bfs}.
\label{fig.3}
}
\end{figure}

Up to now we neglected the contribution of the lightest neutrino mass $m_0=m_{1}$
to $|m_{\beta\beta}|$. From (\ref{eq:32}) it follows that $m_{2}\simeq 0.2\, m_{3}$. 
If we (arbitrarily) assume that $m_{1}\simeq 0.2\, m_{2}$ it follows 
that the modulus of the first term in 
(\ref{eq:37}) is about half of the modulus of the second solar term. Thus, the contribution of the 
$m_{1}$-term to $|m_{\beta\beta}|$ can be sizable. In Fig. \ref{fig.2} we present 
three-dimensional plots for the half-life of the $0\nu\beta\beta$-decay of $^{76}Ge$
as a function of $m_{1}$ and $\sin^{2}\theta_{13}$ under different assumptions 
for the relative CP Majorana phases. In the case of other isotopes similar results are expected.
From  Fig. \ref{fig.2} it follows that for a small values of 
$\sin^{2}\theta_{13}\le 0.01$ the pronounced
hill region with relatively large values of the half-lives 
is not more accessible.

In the case of the inverted hierarchy of neutrino masses the upper and lower bounds on the 
effective Majorana mass $|m_{\beta\beta}|$ are well determined by the known oscillation parameters 
(see (\ref{eq:28})). In Table \ref{tab.3} we give the corresponding ranges for the 
half-lives of relevant nuclei calculated with the nuclear matrix elements of \cite{rodin2} 
by taking into account the best fit and $90\%~C.L.$ values of the parameters 
$\sin^{2}\theta_{12}$ and  $|\Delta m^{2}_{13}|$. In the same Table the predicted ranges of
the half-lives are compared to the current experimental lower limits
on the $0\nu\beta\beta$-decay half-life from experiments with saturated sensitivity 
and from two running experiments (NEMO 3 \cite{nemo3} and Cuoricino \cite{cino})
together with half-lives sensitivities of proposed future experiments.
We note that the designed future
$0\nu\beta\beta$-decay experiments allow different strategies. 
Some of them plan to proceed with smaller steps forward and some of them
prefer to make large steps. The time-scale for these experiments except the GERDA I \cite{gerda}, 
which main task is to confirm or rule out the recent claim
of evidence for the $0\nu\beta\beta$-decay of $^{76}Ge$ \cite{claim}, 
is not determined yet. From Table \ref{tab.3} we conclude that the
sensitivities of the next generation   $0\nu\beta\beta$-decay experiments 
will apparently allow to probe the inverted hierarchy of neutrino masses. 
Finally, in Fig.\ref{fig.3} we present the expected life-times of the 
$0\nu\beta\beta$-decay of $^{76} \rm{Ge}$, $^{130} \rm{Te}$,
$^{136} \rm{Xe}$ and other nuclei calculated for  90\% CL allowed
values of the parameters. The dependence of the $0\nu\beta\beta$-decay 
half-life on $\sin^2\theta_{12}$ is outlined.

\section{Conclusion}

The effective Majorana mass, which determines the half-life of neutrinoless double $\beta$-decay,
crucially depends on the 
character of the neutrino mass spectrum. All possible physical neutrino mass spectra
(hierarchy, inverted hierarchy and quasi degenerate) are at present viable. However, in the framework 
of the see-saw mechanism, which is a plausible explanation of the smallness of neutrino masses,  
hierarchy is a favorable spectrum (see, \cite{Altarelli,Feruglio}). Generically, the 
 neutrino mass hierarchy 
naturally appears in GUT models like SO(10) which unify quarks, leptons and neutrinos. 

Having in mind that the hierarchy of neutrino masses is a plausible neutrino mass spectrum,
in this case
we performed detailed calculations of  half-lives of the $0\nu\beta\beta$-decay of several nuclei of
experimental interest. We used nuclear matrix elements which were obtained in the recent 
most updated and most comprehensive RQRPA calculations \cite{rodin2}.
We studied the dependence of the half-lives of $0\nu\beta\beta$-decay on 
$\sin^2\theta_{13}$ and the lightest neutrino mass
$m_1$. The current \cite{CHOOZ} limit and future \cite{Chooz2,T2K}
sensitivities to the value of the parameter  $\sin^2\theta_{13}$  were considered. 
The calculated lower
limits on the half-lives of nuclei considered are listed in Table \ref{tab.2}. As it is seen they
are in the range  of $10^{28}-10^{30}$ years. 
The expected half-life sensitivities of the next generation of the $0\nu\beta\beta$-decay 
experiments are significantly lower.

The future 
$0\nu\beta\beta$-decay experiments will probe the inverted hierarchy 
of neutrino masses which requires a symmetry in the neutrino mass matrix.
Using the updated  RQRPA calculations of NME \cite{rodin2} we have calculated half-lives 
of $0\nu\beta\beta$-decay of several nuclei in this  case.

\acknowledgments

The work of F. \v{S}. was supported in part by the Deutsche
Forschungsgemeinschaft (436 SLK 17/298 and TU 7/134-1). We thank also  
the EU ILIAS project under the contract RII3-CT-2004-506222. 
S.M.B. acknowledges the support of  the Italien Program  ``Rientro dei cervelli''.

\end{document}